\documentclass[10pt,preprint]{sigplanconf}

\usepackage[normalem]{ulem}
\usepackage{amsmath}
\usepackage{amsmath, amsfonts}
\usepackage{color}
\usepackage{algorithm}
\usepackage{algorithmic}
\usepackage{multirow}
\usepackage{bm}
\usepackage{hyperref}
\usepackage{graphicx}
\usepackage{listings}
\newcommand{\vect}[1]{\boldsymbol{\mathbf{#1}}}
\newtheorem{problem}{Problem}

\newcommand{\ie}{\emph{i.e.}}
\newcommand{\eg}{\emph{e.g.}}
\newcommand{\treename}{$W$-ary}

\newif\ifdebug
\debugfalse

\ifdebug
\newcommand{\chris}[1]{{\color{red}{\bf\sf [CJF: #1]}}}
\newcommand{\junz}[1]{{\color{blue}{\bf\sf [ZJ: #1]}}}
\newcommand{\lkw}[1]{{\color{cyan}{\bf\sf [LKW: #1]}}}
\newcommand{\cwg}[1]{{\color{green}{\bf\sf [CWG: #1]}}}
\newcommand{\hct}[1]{{\color{green}{\bf\sf [HCT: #1]}}}

\else
\newcommand{\chris}[1]{{\color{red}{}}}
\newcommand{\junz}[1]{{\color{blue}{}}}
\newcommand{\lkw}[1]{{\color{cyan}{}}}
\newcommand{\cwg}[1]{{\color{green}{}}}
\newcommand{\hct}[1]{{\color{green}}}
\fi

\begin{document}

\title{SaberLDA: Sparsity-Aware Learning of Topic Models on GPUs}
\date{}

\authorinfo{Kaiwei Li \and Jianfei Chen \and Wenguang Chen \and Jun Zhu}
          {Tsinghua Universtiy}
          {\{likw14, chenjian14\}@mails.tsinghua.edu.cn \and \{cwg, dcszj\}@tsinghua.edu.cn}
\maketitle

\thispagestyle{empty}

\begin{abstract}
Latent Dirichlet Allocation (LDA) is a popular tool for analyzing
discrete count data such as text and images. Applications
require LDA to handle both large datasets and a large
number of topics. Though distributed CPU systems have
been used, GPU-based systems have emerged as a promising
alternative because of the high computational power and
memory bandwidth of GPUs. However, existing GPU-based
LDA systems cannot support a large number of topics because
they use algorithms on dense data structures whose time and
space complexity is linear to the number of topics.

In this paper, we propose SaberLDA, a GPU-based LDA
system that implements a sparsity-aware algorithm to achieve
sublinear time complexity and scales well to learn a large
number of topics. To address the challenges introduced by
sparsity, we propose a novel data layout, a new warp-based
sampling kernel, and an efficient sparse count matrix updating
algorithm that improves locality, makes efficient utilization
of GPU warps, and reduces memory consumption. Experiments
show that SaberLDA can learn from billions-token-scale
data with up to 10,000 topics, which is almost two orders
of magnitude larger than that of the previous GPU-based
systems. With a single GPU card, SaberLDA is able to learn
10,000 topics from a dataset of billions of tokens in a few
hours, which is only achievable with clusters with tens of machines
before.
\end{abstract}

%

\section{Introduction}

Topic models provide a suite of widely adopted statistical tools for feature extraction and dimensionality reduction for bag-of-words (\ie, discrete count) data, such as text documents and images in a bag-of-words format ~\cite{cao2007spatially}. 
Given an input corpus, topic models automatically extract a number of latent \emph{topics}, which are unigram distributions over the words in a given vocabulary. The high-probability words in each topic are semantically correlated.
Latent Dirichlet Allocation~ \cite{blei2003latent} (LDA) is the most popular of topic models due to its simplicity, and has been deployed as a key component in data visualization~\cite{iwata2008probabilistic}, 
text analysis~\cite{boyd2007topic,zhu2009medlda}, computer vision~\cite{cao2007spatially},  network analysis~\cite{Chang:RTM09}, and recommendation systems~ \cite{chen2009collaborative}. 

In practice, it is not uncommon to encounter large-scale datasets,  \eg, text analysis typically consists of hundreds of millions of documents~ \cite{yuan2014lightlda}, and recommendation systems need to tackle hundreds of millions of users~ \cite{ahmed2012scalable}. Furthermore, as the scale of the datasets increases, the model size needs to be increased as well --- we need a larger number of topics in order to exploit the richer semantic structure underlying the data. 
A reasonable design goal for modern topic modeling systems is thousands of topics, to have a good coverage for both industry scale applications~\cite{wang2014peacock} and researching~\cite{boyd2007topic,iwata2008probabilistic, cao2007spatially}.

However, it is highly challenging to efficiently train large LDA models. The time complexity of training LDA is high because it involves iteratively scanning the input corpus for many times (\eg, 100), and the time complexity of processing each token is not constant but related to the number of topics.

\begin{table}[t]
\caption{A summary of GPU-based LDA systems, $D$: the number of documents, $K$: the number of topics, $V$: the size of vocabulary, $T$: the number of tokens. \chris{Add model size and memory size, focus on larger $K$ than GPU.}
\label{tbl:previous-systems}}
\centering
\begin{tabular}{lllll}
\hline
 Implementation           & $D$  & $K$ & $V$ & $T$ \\ \hline
Yan et al.~ \cite{yan2009parallel} &  300K & 128 & 100K & 100M \\ 
BIDMach~ \cite{zhao2015same} &  300K  & 256 & 100K & 100M \\
Steele and Tristan~ \cite{tristan2015efficient}  & 50K  & 20 & 40K & 3M  \\
\textbf{SaberLDA}                  & 19.4M   & \textbf{10K} & 100K & 7.1B \\
\hline
\end{tabular}
\end{table}

To train LDA in acceptable time, CPU clusters are often used. However, due to the limited memory bandwidth and low computational power of CPUs, large clusters are typically required to learn large topic models~\cite{ahmed2012scalable,yuan2014lightlda,yu2015scalable}. For example, a 32-machine cluster is used to learn 1,000 topics from a 1.5-billion-token corpus.

A promising alternative is to train LDA with graphics processing units (GPUs), leveraging their high computational power and memory bandwidth. Along this line, there have been a number of previous attempts. For example, Yan et al.~\cite{yan2009parallel} implement the collapsed Gibbs sampling algorithm, BIDMach~\cite{zhao2015same} implements the variational Bayes algorithm as well as a modified Gibbs sampling algorithm, and Steele and Tristan~\cite{tristan2015efficient} propose a expectation-maximization algorithm. These GPU-based systems are reported to achieve superior performance than CPU-based systems~\cite{yan2009parallel,zhao2015same,tristan2015efficient}.

Unfortunately, current GPU-based LDA systems can only learn a few hundred topics (See Table~\ref{tbl:previous-systems}), which may not be sufficient to capture the rich semantic structure underlying the large datasets in industry scale applications~\cite{wang2014peacock}. It is fundamentally difficult for these systems to learn more topics because they use algorithms on dense data structures whose time and space complexity is linear to the number of topics.


To address this problem, we propose SaberLDA, a novel GPU-based system that adopts a \emph{sparsity-aware algorithm} for LDA. Sparsity aware algorithms are based on the insight that a single document is not likely to have many topics, and are able to achieve \emph{sub-linear} (or even amortized constant) time complexity with respect to the number of topics. Representative examples include AliasLDA~\cite{li2014reducing}, F+LDA~\cite{yu2015scalable}, LightLDA~\cite{yuan2014lightlda}, WarpLDA~\cite{chen2016warplda} and ESCA~\cite{zaheer2015exponential}, which are implemented in general purpose CPU systems. Therefore, the running time is not sensitive with the number of topics.  

However, it is considerably more challenging to design and
implement sparsity-aware algorithms on GPUs than on CPUs.
Comparing with CPUs, GPUs have much larger number of
threads, much smaller per thread cache size and longer cache
lines, which makes it much more difficult to use caches to mitigate
the random access problems introduced by sparsity. The
branch divergence issues of GPUs suggest we should fully
vectorize the code. But for sparse data structures, loop length
is no longer fixed and the data are not aligned, which indicates
that straightforward vectorization is not feasible. Finally, the
limited GPU memory capacity requires streaming input data
and model data, which add another dimension of complexity
for data partition and layout, to enable parallelism, good
locality and efficient sparse matrix updates all together.


SaberLDA addresses all these challenges for supporting sparsity-aware algorithms on GPU, our key technical contribution includes:
\begin{itemize}
\item A novel hybrid data layout ``partition-by-document and order-by-word'' (PDOW) that simultaneously maximizes the locality and reduces the GPU memory consumption;
\item A warp-based sampling kernel that is fully vectorized, and is equipped with a \treename sampling tree that supports both efficient construction and sampling;
\item An efficient ``shuffle and segmented count''  (SSC) algorithm for updating sparse count matrices.
\end{itemize}

Our experimental results demonstrate that SaberLDA is able to train LDA models with up to 10,000 topics, which is more than an order of magnitude larger than previous GPU-based systems~\cite{yan2009parallel,zhao2015same,tristan2015efficient}, where the throughput only decreases by 17\% when the number of topics increases from 1,000 to 10,000. SaberLDA is also highly efficient, under various topic settings, SaberLDA converges 5 times faster than previous GPU-based systems and 4 times faster than CPU-based systems. With a single card, SaberLDA is able to learn 10,000 topics from a dataset of billions of tokens, which is only achievable with clusters with tens of machines before~\cite{yu2015scalable}.

The rest the paper is organized as follows, Sec.~\ref{sec:basics} introduces the basics for LDA. Sec.~\ref{sec:design-and-implementation} presents the design of SaberLDA. Sec.~\ref{sec:eval} contains the experiments and Sec.~\ref{sec:conclusions} concludes.

\chris{TODO: examine all figure captions}

\section{Latent Dirichlet Allocation}\label{sec:basics}
In this section, we introduce the latent Dirichlet allocation (LDA) model and its sampling algorithm.
\subsection{Definition}

\begin{figure}
\centering
\includegraphics[width=\linewidth]{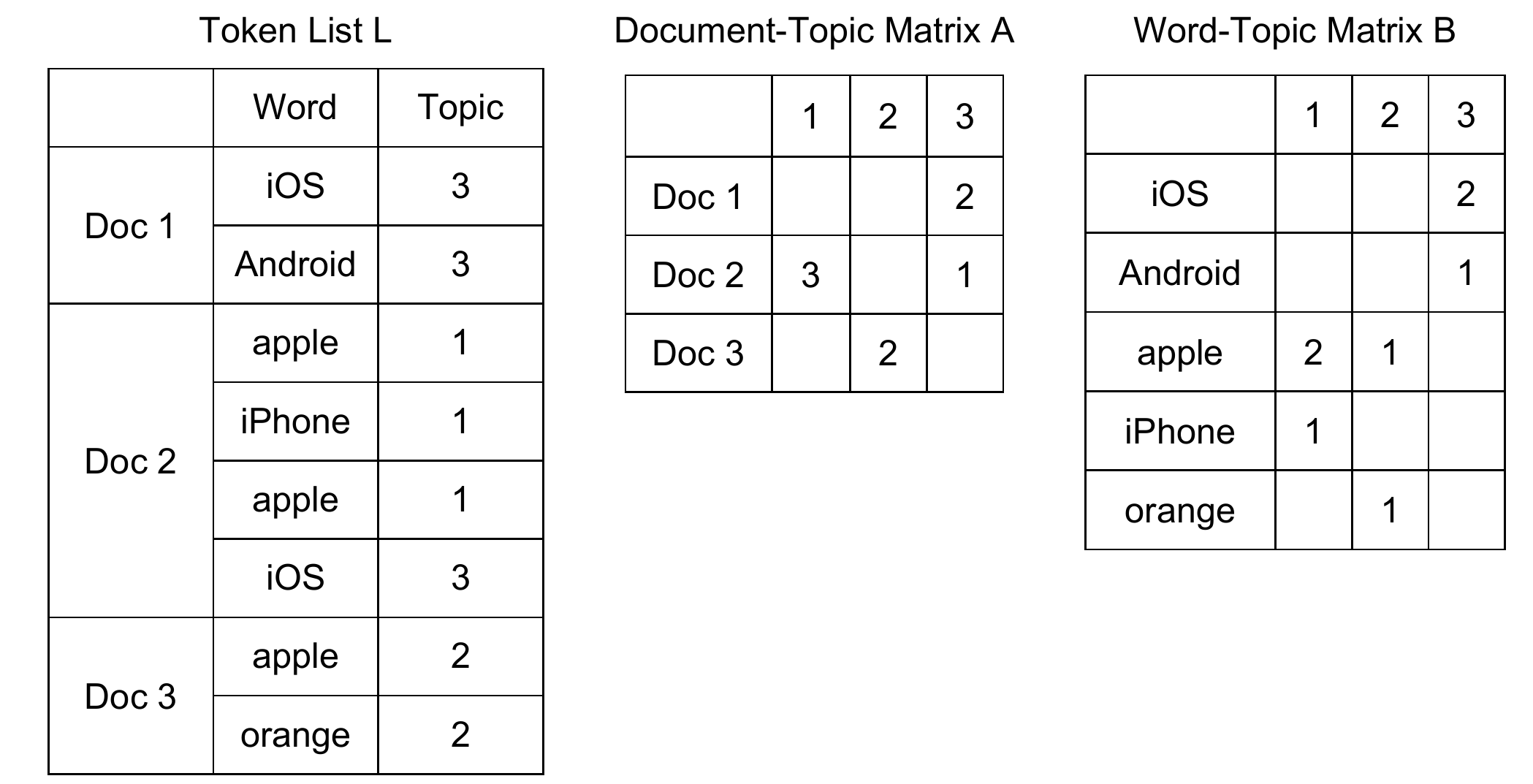}
\caption{An example token list and related count matrices.}
\label{fig:lda-intro}
\end{figure}

LDA is a hierarchical Bayesian model that learns latent topics from text corpora~\cite{blei2003latent}.
For an input text corpus, the scale of the learning task is determined by the following four numbers:
\begin{itemize}
\item $D$: the number of documents in the given corpus;
\item $T$: the number of tokens, \ie,  words, appearing in all the documents;
\item $V$: the number of unique words in the corpus, also known as vocabulary size;
\item $K$: the number of topics,
\end{itemize}
where $D$, $T$ and $V$ are determined by the corpus, and $K$ is a parameter that users can specify.

The text corpus is represented as a \emph{token list} $L$, where each occurrence of word $v\in [1, V]$ in document $d\in [1, D]$ is called a \emph{token}, and represented by a triplet $(d, v, k)$. 
While $d$ and $v$ are specified by the corpus, training LDA involves assigning a \emph{topic assignment} $k\in [1, K]$ for each token $(d, v, k)$.  

After the token list is given, we construct two count matrices as below:
\begin{itemize}
\item The document-topic count matrix $\vect A$ is $D\times K$, where $A_{dk}$ is the number of tokens $t$ with $(d_t=d, k_t=k)$. 
\item The word-topic count matrix $\vect B$ is $V\times K$, where $B_{vk}$ is the number of tokens $t$ with $(v_t=v, k_t=k)$,
\end{itemize}
where the matrix $\vect A$ is often sparse, \ie, has many zero elements (See below for an example). 

We omit the mathematical details of LDA and refer the interested readers to standard LDA literature~\cite{blei2003latent,griffiths2004finding}. Informally, the LDA model is designed in a way that maximizes some objective function (the likelihood) related to the topic assignments $k$. A document $d$ can be characterized by the $d$-th row of the document-topic matrix $\vect A$, and a topic $k$ can be characterized by the $k$-th column of the word-topic matrix $\vect B$. Figure~\ref{fig:lda-intro} is a concrete example, where there are $D=3$ documents, $T=8$ tokens, $V=5$ words in the vocabulary, and $K=3$ topics. Each token is assigned with a topic assignment from 1 to 3, indicated as the superscript in the figure. $\vect A$ is the document-topic matrix, \eg, $A_{13}=2$ because both tokens in document 1 is assigned to topic 3. $\vect B$ is the word-topic matrix, and its each column characterizes what the corresponding topic is about, \eg, the first topic has the word ``apple'' and ``iPhone'', and is about the device iPhone; the second topic is about fruits, and the third topic is about mobile OS. Likewise, the rows of $\vect A$ characterizes what documents are about, \eg, the first document is about mobile OS, the second document is about iPhone and mobile OS, and the last document is about fruits. Note that the matrix $\vect A$ is sparse, because a document is not likely to be relevant with all the topics at the same time. 

\subsection{Inference}\label{sec:inference}
Given the token list $L$ , our goal is to infer the topic assignment $k_t$ for each token $t$. Many algorithms exist for this purpose, such as variational inference~\cite{blei2003latent}, Markov chain Monte-Carlo~\cite{griffiths2004finding}, and expectation maximization~\cite{zaheer2015exponential, chen2016warplda}. We choose the ESCA algorithm~\cite{zaheer2015exponential} to implement on GPU for its following advantages
\begin{itemize}
\item It is sparsity-aware, so the time complexity is sub-linear with respect to the number of topics. This property is critical to support the efficient training of large models;
\item It enjoys the best degree of parallelism because the count matrices $\vect A$ and $\vect B$ only need to be updated once per iteration. This matches with the massively parallel nature of GPU to achieve high performance.
\end{itemize}
The ESCA algorithm alternatively performs the E-step and the M-step for a given number of iterations (e.g., 100 iterations), where the E-step updates $k_t$ given the counts $\vect A, \vect B$, and the M step updates $\vect A, \vect B$ given $k_t$:
\begin{itemize}
\item E-step: for each token $(d, v, k)$, sample the topic assignment $k$:
\begin{align}\label{eqn:lda}
p(k)\propto (A_{dk}+\alpha)\hat B_{vk},
\end{align}
where $\propto$ means ``proportional to'' and the word-topic probability matrix $\vect {\hat B}$ is a normalized version of the count matrix $\vect B$ in the sense that $\vect {\hat B}$ is roughly proportional to $\vect B$, but each of its column sums up to 1. $\hat B_{vk}$ is computed as follows:
\begin{align}\label{eqn:preprocess}
\hat B_{vk}=\frac{B_{vk}+\beta}{\sum_{v=1}^V B_{vk} + V\beta},
\end{align}
\item M-step: update the counts $\vect A$ and $\vect B$, and calculate $\vect {\hat B}$.
\end{itemize}
Here, $\alpha$ and $\beta$ are two user specified parameters that control the granularity of topics. Large $\alpha$ and $\beta$ values mean that we want to discover a few general topics, while small $\alpha$ and $\beta$ values mean that we want to discover many specific topics. 

The updates Eq.~(\ref{eqn:lda},~\ref{eqn:preprocess}) can be understood intuitively. A token $(d, v, k)$ is likely to have the topic assignment $k$ if both $A_{dk}$ and $B_{vk}$ are large, \ie, a lot of tokens in document $d$ are topic $k$ and a lot of tokens of word $v$ are topic $k$. For example, if we wish to update the topic assignment of the ``apple'' in document 3, it is more likely to be topic 2 rather than topic 1, because the other token ``orange'' in the same document is assigned to topic 2.


\subsection{Sampling from a multinomial distribution}\label{sec:sampling-multinomial}

\newcommand{\naive}{na\"{i}ve}

The core of the above algorithm is sampling according to
Eq.~(\ref{eqn:lda}). To help readers understand the sampling procedure,
we begin with a vanilla sampling algorithm, and then proceed
to a more complicated sparsity-aware algorithm. Sampling
according to Eq.~(\ref{eqn:lda}) can be viewed as throwing a needle onto
the ground, and report the number of the region where the needle
falls in, where the area of region $k$ is the probability $p(k)$
(Figure~\ref{fig:needle}).
This can be implemented with three steps:

\begin{figure}[t]
\centering
\includegraphics[width=0.9\linewidth]{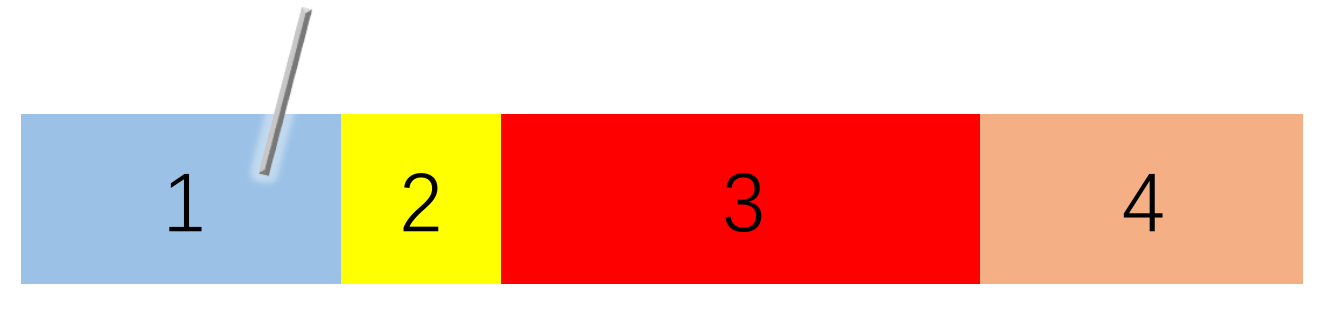}
\caption{Sampling from a multinomial distribution, areas are
proportional to the probabilities, $p(k=1)=0.25, p(k=2)=0.125, p(k=3)=0.375, p(k=4)=0.25$.\label{fig:needle}}
\end{figure}

\begin{enumerate}
\item For $k=1, \dots, K$, compute the probabilities $p(k)$ and their sum $S=\sum_k p(k)$;
\item Generate a random number $u\in [0, S)$;
\item Compute the prefix sum  $c_k = c_{k-1} + p(k)$, where $c_0 = 0$, and return the first $k$ such that $u\le c_k$, with a binary search.
\end{enumerate}
We refer the above procedure as the vanilla algorithm, whose time complexity is limited by step 1 and step 3, which are $O(K)$. Step 3 is a very important routine which we will use over and over again, and we refer its result as ``the position of $u$ in the prefix sum array of $p(k)$'' for brief.

While the vanilla algorithm is $O(K)$, the sparsity-aware algorithms~\cite{li2014reducing,zaheer2015exponential} utilize the sparsity of $\vect A$, and improve the time complexity to $O(K_d)$, where $K_d$ is the average number of non-zero entries per row of $\vect A$. The algorithm decomposes the original sampling problem as two easier sampling sub-problems. For sampling each token, it returns the result of a random  sub-problem, where the probability of choosing the first sub-problem is $\frac{S}{S+Q_v}$, where $S=\sum_{k=1}^K A_{dk} \hat B_{vk}$ and $Q_v=\alpha \sum_{k=1}^K \hat B_{vk}$. The two sub-problems are defined as follows:
\begin{problem}
Sample $p_1(k)\propto A_{dk} \hat B_{vk}$.
\end{problem}
This can be sampled with the vanilla algorithm we described before. But \emph{sparsity} can be utilized: if $A_{dk}=0$, then $p_1(k)=0$ as well. Therefore, we only need to consider the indexes $k$ where $A_{dk}\ne 0$. There are only $K_d$ such indexes on average, therefore, the time complexity is only $O(K_d)$ instead of $O(K)$. Similarly, $S$ can be computed in $O(K_d)$.

\begin{problem}
Sample $p_2(k)\propto \alpha \hat B_{vk} \propto \hat B_{vk}$.
\end{problem}
This problem is only relevant with $v$ but not $d$. We can pre-process for each $v$. There are various approaches for pre-processing which we will cover in detail in Sec.~\ref{sec:warp-based-sampling}. In brief, we construct a tree $T_v$ for each $v$, and then each sample can be obtained with the tree in $O(\log_W K)$. The pre-processing step is not the bottleneck because it is done only once per iteration. 

\begin{algorithm}[t]
  \caption{ESCA algorithm for LDA.}
  \label{algo:lda}
\begin{algorithmic}[1]
    \STATE \textbf{Input:} token list $L$
    \STATE \textbf{Variable:} sparse matrix $\vect A$, dense matrix $\vect{B}$ and $\vect{{\hat B}}$, vector $Q$, sampling trees of all unique words $\vect T$

    \FOR {$i\leftarrow 1 \mbox{ to } \mbox{num\_iteration}$}
    \STATE// E Step:
    \FOR {$(d, v, k) \in L$ }
    \STATE  $k \leftarrow \mbox{Sample}(\vect A_{d}, \vect {\hat B}_{v}, Q_v, T_v$)
    \ENDFOR
    \STATE// M Step:
    \STATE $\vect A \leftarrow \mbox{CountByDZ}(L)$
    \STATE $\vect B \leftarrow \mbox{CountByVZ}(L)$ 
    \STATE $\vect {\hat B}\leftarrow \mbox{Preprocess}(\vect B, \beta)$
    \FOR {$v \leftarrow 1 \mbox{ \textbf{to} } V$}
        \STATE $Q_v, T_v\leftarrow \mbox{BuildTree}(\vect{\hat{B}}_v)$
    \ENDFOR
\ENDFOR
\end{algorithmic}
\end{algorithm}

\subsection{Pseudo-Code}\label{sec:pseudo-code}
To make the presentation more concrete, we present the pseudocode of the ESCA algorithm as Alg.~\ref{algo:lda}, which is a Bulk Synchronous Parallel (BSP) programming model. 
In the E-step of each iteration, all the tokens in the token list $L$ are updated independently, by calling \textbf{Sample} for each token. All arguments of the function \textbf{Sample} are read-only, and the return value is the new topic assignment $k$.
In the M-step, the matrices $\vect A$ and $\vect B$ are calculated from the token list $L$, by functions \mbox{CountByDZ} and \mbox{CountByVZ}.
Then, $\vect {\hat B}$ is updated by the function \textbf{Preprocess} following Eq.~(\ref{eqn:preprocess}). Finally, generate the sampling trees $T_v$ and sums $Q_v$ for each word $v$.

\begin{algorithm}[t]
 \caption{Sparsity aware sampling}
  \label{algo:kernel}
\begin{algorithmic}[1]
\STATE \textbf{Input:} sparse vector $A$, dense vector $\hat B$, scalar $Q$, tree $T$
\STATE $S \leftarrow 0$
\STATE $P \leftarrow \mbox{new sparse vector}$
\FOR { $k\leftarrow \mbox{non-zero elements of } A$}
\STATE $P_k \leftarrow A_k \times \hat B_{k}$
\STATE $S \leftarrow S +P_k$
\ENDFOR
\IF {$\mbox{random}(0,1) < S / (S + Q)$}
\STATE $k \leftarrow $ sample from $P$
\ELSE
\STATE $k \leftarrow $ $T$.sample()
\ENDIF
\RETURN $k$
\end{algorithmic}
\end{algorithm}


As shown in Alg.~\ref{algo:kernel}, the function \textbf{Sample} samples the topic assignment $k$ given the rows $\vect A_{d}$, $\vect {\hat B}_{v}$, the sum $Q_v$, and the tree $T_v$, by implementing the sparsity-aware algorithm described in Sec.~\ref{sec:sampling-multinomial}. First, we compute the probability for Problem 1, $P=\vect A_{d}\bigodot\vect {\hat B}_{v}$ as well as $S$, where $\bigodot$ is the element-wise product. Next, we flip a coin with the probability of head being $\frac{S}{S + Q}$. If the coin is head, perform sampling from $p_1(k)\propto P_k$, which involves finding the location of a random number in the prefix-sum array of the sparse vector $P$ as discussed in Sec.~\ref{sec:sampling-multinomial}. Otherwise, sample from $p_2(k)\propto \hat B_{vk}$ with the tree $T$, which is pre-processed.

 
 
 

\section{Design and Implementation}\label{sec:design-and-implementation}

In this section we present SaberLDA, a high performance GPU-based system for training LDA. The design goals of SaberLDA are:
\begin{itemize}
    \item supporting large models up to 10,000 topics; 
    \item supporting large dataset of billions of tokens;
    \item providing comparable performance with a moderate size CPU cluster using a single GPU card.
\end{itemize} 
It is difficult for previous GPU-based systems~\cite{yan2009parallel,zhao2015same,tristan2015efficient} to satisfy all these goals, because all of them are based on the vanilla algorithm mentioned in Sec.~\ref{sec:sampling-multinomial} which have  $O(K)$ time complexity, \ie, the training time will be 100 times longer if the number of topics increase from hundreds (current) to 10,000 (goal), which is not acceptable. 

To address this problem, SaberLDA adopts the sparsity-aware sampling algorithm whose time complexity is $O(K_d)$, which is not very sensitive to the number of topics. However, exploiting sparsity is highly challenging on GPUs. The memory access is no longer sequential as that for the dense case, and the small cache size and long cache line of GPU aggravate this problem. Therefore, the memory accesses need to be organized very carefully for good locality.  Moreover, vectorization is not as straightforward as the dense case since the loop length is no longer fixed and the data are not aligned, so waiting and unconcealed memory accesses can happen.\chris{Is it precise?} Finally, updating the sparse count matrix is more complicated than dense count matrices. We now  present our design of SaberLDA which addresses all the challenges.

\subsection{Streaming Workflow\chris{choose a proper title}}\label{sec:data-layout}

We first consider how to implement Alg.~\ref{algo:lda},  deferring the details of the functions \textbf{Sample}, \textbf{CountByDZ}, \textbf{CountByVZ}, \textbf{Preprocess} and \textbf{BuildTree} to subsequent sections. The
implementation involves designing the storage format, data placement (host / device), and partitioning strategies for all the data items\chris{data structures might be better}, as well as designing the order of sampling tokens (Line 5 of Alg.~\ref{algo:lda}). The data items include the token list $L$, the document-topic count matrix $\vect A$, the word-topic count matrix $\vect B$, and the word-topic distribution matrix $\vect {\hat B}$. 

The aforementioned design should maximize the locality, and meanwhile, keep the size of the data items within the budget of GPU memory even when both $K$ and the data size are large. This is challenging because locality emerges as a new design goal to support sparsity-aware algorithms, while previous dense matrix based GPU systems~\cite{yan2009parallel,zhao2015same,tristan2015efficient} only have one goal (\ie, fitting into the memory budget) instead of the two which need to be obtained simultaneously in our method. 
As we will see soon, the simple data layout in previous systems such as sorting all tokens by document-id~\cite{yan2009parallel,zhao2015same,tristan2015efficient,zaheer2015exponential} or by word-id~\cite{yu2015scalable} cannot satisfy both requirements, and we will propose a hybrid data layout to address this problem. 

\subsubsection{Storage Format}

We analyze the access pattern of these data items described in Alg.~\ref{algo:lda} and Alg.~\ref{algo:kernel}. The token list $L$ is accessed sequentially, and we store it with an array. The document-topic count matrix $\vect A$ is accessed by row, and the \textbf{Sample} function iterates over its non-zero entries, so we store it by the compressed sparse rows (CSR) format to avoid enumerating over zero entries.\lkw{The number of non-zero entries can be no more than number of tokens. When k is small (i.e. 100), usually all entries are non-zero, CSR is more memory cost than dense matrix. But when k grows larger, there are more zero entries, hence CSR is greatly memory efficient.} 
Besides efficiency, the CSR format also reduces the memory consumption and host/device data transferring time comparing with the dense matrix format in previous implementations~\cite{yan2009parallel,zhao2015same,tristan2015efficient}.
The word-topic matrices $\vect B$ and $\vect {\hat B}$ are randomly accessed, and we store them as dense matrices. Table~\ref{tab:mem} lists the memory consumption of each data item in the PubMed dataset (See Sec.~\ref{sec:eval} for the details), where we can observe that  representing $\vect A$ as sparse matrix indeed saves a lot of memory when $K\ge 1,000$, and the GPU memory is large enough to hold $\vect B$ and $\vect {\hat B}$ for the desired number of topics. 

\begin{table}[t]
\centering
\caption{Memory Consumption of PubMed Dataset\chris{TODO: measure what is $<$5.8GB, change 10k to 5k? Maybe 10k is okay here? Why dense is 9GB, 38GB, and 330GB?}}
\label{tab:mem}
\begin{tabular}{crrrr}
\hline
\multirow{2}{*}{Data}   & \multicolumn{1}{c}{\multirow{2}{*}{\begin{tabular}[c]{@{}c@{}}Word-Topic\\ Matrix $\vect B$, $\vect {\hat B}$\end{tabular}}} & \multicolumn{1}{c}{\multirow{2}{*}{Token List $L$}} & \multicolumn{2}{c}{Doc-Topic Matrix $\vect A$}                                         \\ \cline{4-5} 
                        & \multicolumn{1}{c}{}                                                                          & \multicolumn{1}{c}{}                            & \multicolumn{1}{c}{Dense}          & \multicolumn{1}{c}{Sparse}           \\ \hline
K                       & V=141k                                                                                        & T=738M                                          & \multicolumn{2}{c}{D=8.2M}                                                \\ \hline
\multicolumn{1}{r}{100} & 0.108 GB                                                                                      & 8.65 GB                                          & 3.2 GB                         & 5.8 GB                      \\
\multicolumn{1}{r}{1k}  & 1.08 GB                                                                                       & 8.65 GB                                          & 32 GB                              & 5.8 GB                      \\
\multicolumn{1}{r}{10k} & 10.8 GB                                                                                        & 8.65 GB                                          & 320 GB                             & 5.8 GB                      \\ \hline
\end{tabular}
\end{table}

\subsubsection{Data Placement and Partitioning Strategy}
\begin{figure}[t]
\centering
\includegraphics[width=\linewidth]{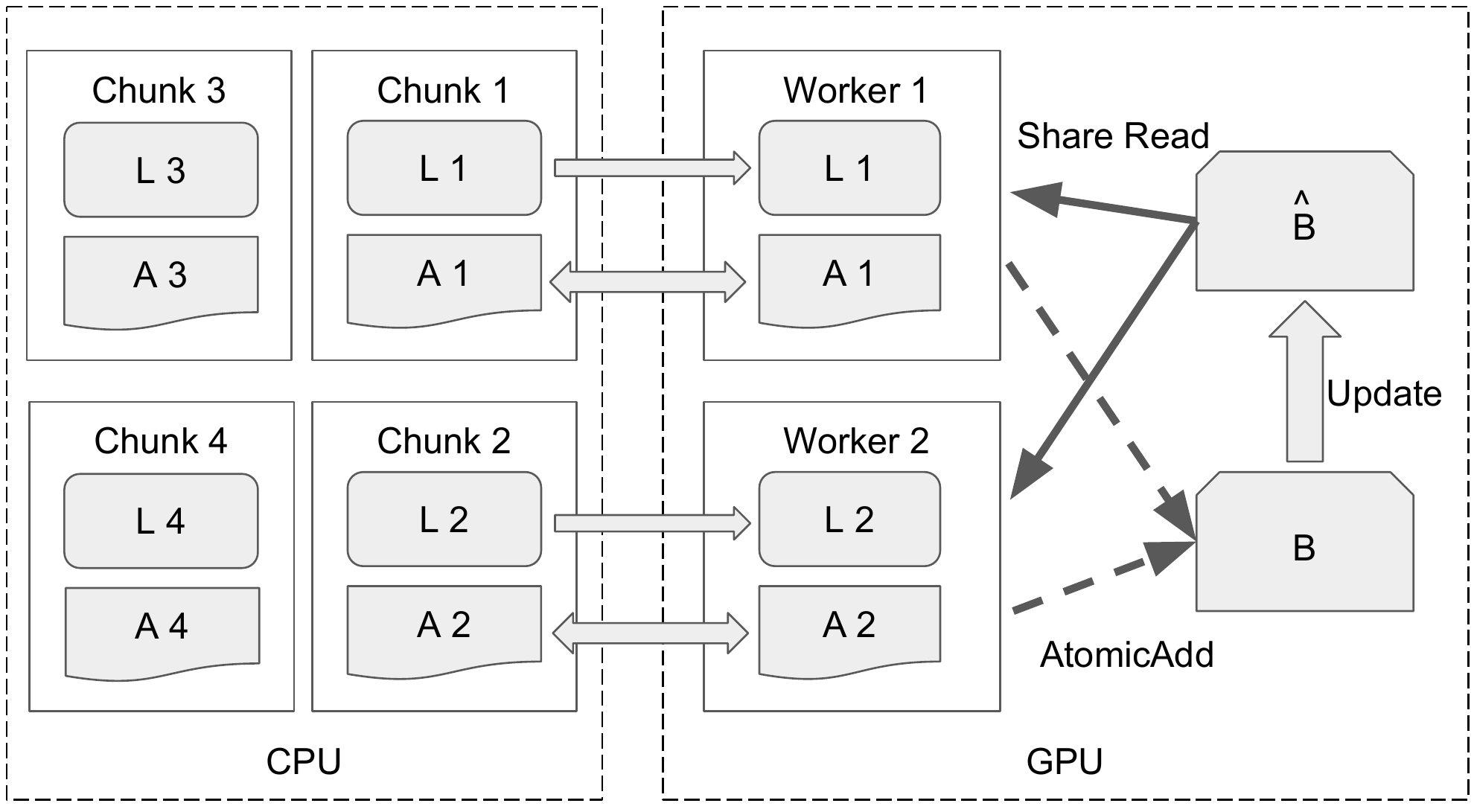}
\caption{Streaming workflow of SaberLDA. $\vect L$: Token List, $\vect A$: document-topic count matrix, $\vect B$: word-topic count matrix, $\vect{\hat{B}}$: word-topic probability matrix.\label{fig:SaberLDA-demo}}
\end{figure}

However, the token list $L$ and the document-topic matrix $\vect A$ cannot fit in the GPU memory, because the $L$ grows linearly with the number of tokens, and the size of $\vect A$ grows linearly with the number of documents. Therefore, the sizes of both data items grow rapidly with the size of the dataset. Since one of our major design goals is to support large datasets with billions of tokens, $L$ and $\vect A$ cannot be held in GPU memory because they can be arbitarily large as the size of the dataset grows. 

To address this problem, we treat $L$ and $\vect A$ as \emph{streams}: we store them in the main memory, partition them into chunks, and process a few chunks with the GPU at a time.\chris{will readers ask whether the main memory is large enough to hold terabyte-scale data?} We partition $L$ and $\vect A$ \emph{by document}, \ie, each chunk has all tokens from a certain set of documents along with the corresponding rows of $\vect A$. Several \emph{workers} are responsible of performing the sampling (Alg.~\ref{algo:kernel}) for the tokens in each chunk, as illustrated in Fig.~\ref{fig:SaberLDA-demo}. Each worker is a \textbf{cudaStream} that fetches a chunk from the main memory, samples the topic assignments for all the tokens in that chunk, updates the document-topic count maxis $\vect A$, and sends the updated $\vect A$ back to the main memory. The computation and communication are overlapped by having multiple workers. The word-topic matrices $\vect B$ and $\vect {\hat B}$ are in the GPU memory, while in the sampling stage, all the workers read $\vect {\hat B}$ and updates $\vect B$. When the sampling of each iteration is finished, $\vect {\hat B}$ is updated with $\vect B$.

\subsubsection{Order of Sampling Tokens}\label{sec:order-of-sampling}
\begin{figure}[t]
\centering
\includegraphics[width=\linewidth]{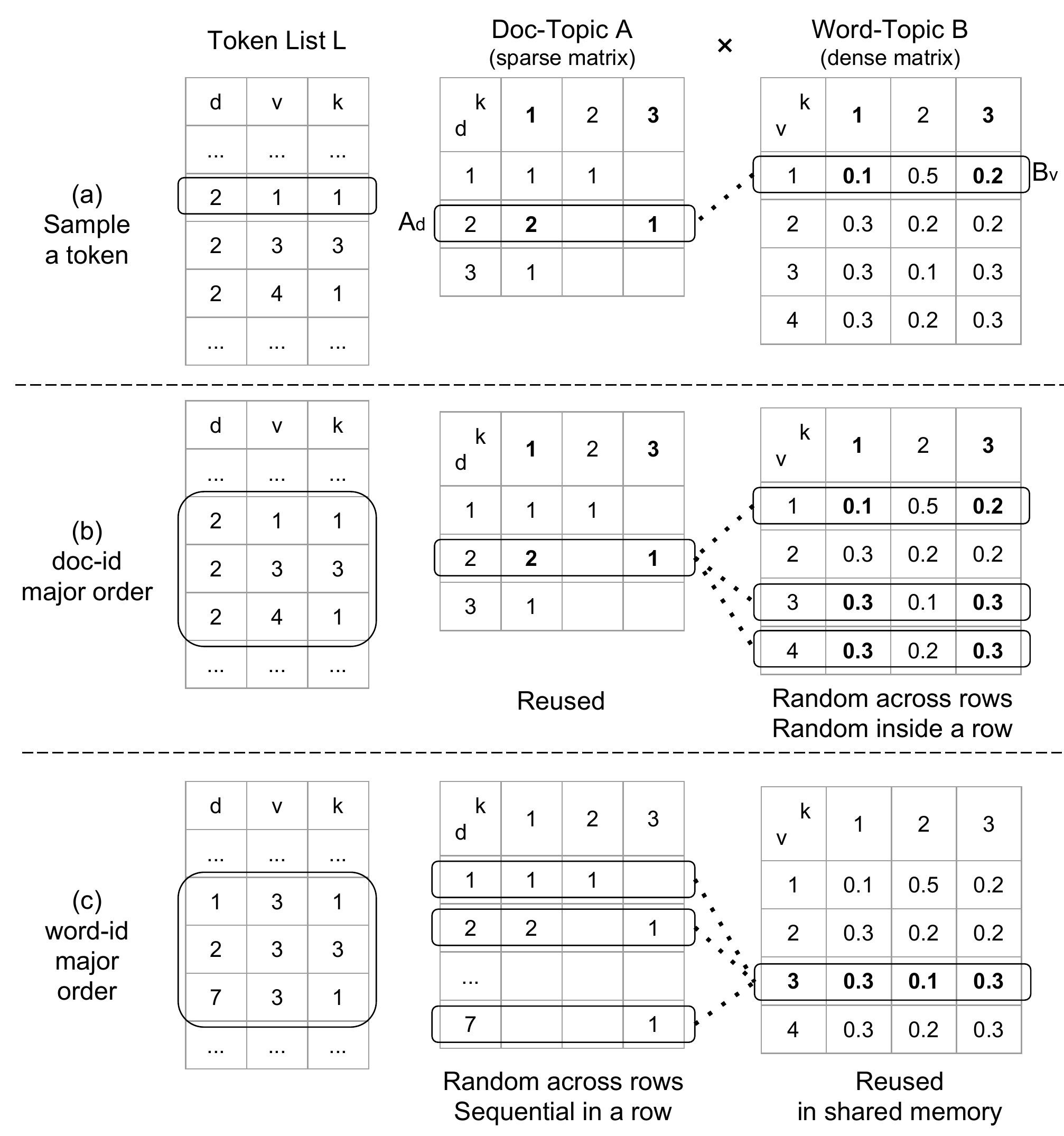}
\caption{Memory Access Pattern of Sampling.\label{fig:locality}}
\end{figure}

We now discuss the order
of sampling tokens, \ie, executing \textbf{Sample} for each token in Alg.~\ref{algo:lda}. Although theoretically these tokens can be sampled in any order, the ordering greatly impacts the locality, as we shall see soon; therefore calling for a careful treatment.


As illustrated as Fig.~\ref{fig:locality}(a), sampling $k$ for a token $(d, v, k)$ requires evaluating a element-wise product of two rows (Line 4-7 of Alg.~\ref{algo:kernel}), \ie, the $d$-th row of the document-topic count matrix $\vect A_{d}$ and the $v$-th row of the word-topic probability matrix $\vect {\hat{B}}_{v}$. The element-wise product  involves accessing all non-zero entries of $\vect A_{d}$ (sequential), and accessing the elements of $\vect {\hat{B}}_{v}$ indexed by the non-zero entries of $\vect A_{d}$ (random). 

There are two particular visiting orders which  reuse the result of previous memory accesses for better locality. The \emph{doc-major order} sorts the tokens by their document-id's, so that the tokens belonging to the same document are adjacent in the token list. Before processing the tokens in the $d$-th document, the row $\vect A_d$ can be fetched into the shared memory and reused for sampling all the subsequent tokens in that document. On the other hand, the access of $\vect {\hat B}_v$ cannot be reused, because each token requires to \emph{access random elements} (indexed by the non-zero entries of $\vect A_d$) \emph{in random row} of the word-topic count matrix $\vect {\hat B}$, as shown in Fig.~\ref{fig:locality} (b). The bottleneck of this approach is accessing $\vect {\hat B}$, where both the row index and column index are random. 

On the contrary, the \emph{word-major order} sorts the tokens by their word-id's, so that the tokens belong to same word are adjacent in the token list. Before processing the tokens of the $v$-th word, the row $\vect {\hat B}_v$ can be fetched into the shared memory and reused for sampling all the subsequent tokens of that word. Each token needs to \emph{access all the elements of a random row} $\vect A_d$ (Fig.~\ref{fig:locality} (c)). The bottleneck of this approach is accessing $\vect A$, where only the row index is random. 

The memory hierarchies are quite different on CPUs and GPUs. CPUs have larger cache size ($>$30MB) and shorter cache line (64B), while GPUs have smaller cache size ($>$2MB) and longer cache line.\chris{Add reference of GPU memory hierarchy} On CPUs, when $K$ and $V$ are small, the document-major order can have better cache locality than the word-major order because $\vect {\hat B}$ can fit in cache~\cite{zaheer2015exponential}. But for GPUs, the word-major order has clear advantage that it efficiently utilizes of the cache line by accessing whole rows of $\vect A_d$ instead of random elements. Therefore, we choose the word-major order for SaberLDA.\chris{I may confuse ``have good locality'' and ``fully utilize the cache line'', is it okay?}

\subsubsection{Partition-by-Document and  Order-by-Word}\label{sec:pdow}
Putting all the above analysis together, SaberLDA adopts a \emph{hybrid} data layout called partition-by-document and order-by-word (PDOW), which means to firstly partition the token list by document-id, and then sort the tokens within each chunk by word-id. Unlike the simple layouts such as sorting all tokens by document-id or by word-id in previous systems~\cite{yan2009parallel,zhao2015same,tristan2015efficient,zaheer2015exponential,yuan2014lightlda,yu2015scalable}, PDOW combines the advantages of both by-document partitioning and word-major ordering, and simultaneously improves cache locality with the word-major order, and keeps the GPU memory consumption small with the by-document partitioning. 

The number of chunks presents a tradeoff between memory consumption and locality. The memory consumption is inversely proportional to the number of chunks, but the number of times to load each row of $\vect B$ into shared memory is proportional to the number of chunks. The chunk size should not be too small to ensure good locality, \ie, the pre-loaded $\vect B_v$ should be reused for a suffciently large number of times. In SaberLDA, we minimize the number of chunks as long as the memory is sufficient.

\chris{This is the most significant technique in our paper. To let readers have a clear picture in their mind, draw a example token list where we partition by document and order by word, as Figure 4(d).}

\subsection{Warp-based Sampling}\label{sec:warp-based-sampling}
\lstdefinestyle{customc}{
  belowcaptionskip=1\baselineskip,
  breaklines=true,
  frame=L,
  xleftmargin=\parindent,
  language=C,
  showstringspaces=false,
  basicstyle=\footnotesize\ttfamily,
    morekeywords={__shared__, thread_id, class, private, public},
}

\lstset{escapechar=@,style=customc}

\begin{figure}
\begin{lstlisting}[]
int WarpSample(SparseVector A, __shared__ DenseVector B_hat, Wary_tree T, RandomSeed &seed) {
    float S = 0;
    __shared__ float P[A.size()];
    for (int i=thread_id; i<A.size(); i+=32) {
        int k = A[i].idx;
        P[i] = A[i].val * B_hat[k];
        S += P[i];
    }
    S = warp_sum(S);
    if (RandomFloat(seed) < S/(S+T.sum())) {
        float x = RandomFloat(seed) * S;
        float ps = 0;
        for (int i=0; i<A.size(); i+=32) {
            ps += warp_prefix_sum(P[i+thread_id]);
            int vote_id = warp_vote(ps >= x);
            if (vote_id != -1)
                return A[i+vote_id].idx;
            ps = warp_copy(ps, 31);
        }
    } else return T.Sample(seed);
}
\end{lstlisting}
\caption{Warp-based sampling kernel. \label{fig:warp-sampling}}
\end{figure}

We now turn to the \textbf{Sample} function (Alg.~\ref{algo:kernel}), which is the most time consuming part of LDA. To understand the challenges and efficiency-related considerations, it is helpful to have a brief review of GPU architecture. 

GPUs follow a single instruction multiple data (SIMD) pattern, where the basic SIMD unit is \emph{warp}, which has 32 \emph{data lanes}. Each lane has its own ALU and registers, and all the lanes in a warp execute the same instruction. In CUDA, each thread is executed on a lane, and every adjacent 32 threads share the same instruction. Readers can make an analogy between GPU warp instruction and CPU vector instruction. 

The most straightforward implementation of sampling on GPU is \emph{thread-based sampling}, which samples each token with a GPU thread. Therefore, 32 tokens are processed in parallel with a warp. Thread-based sampling is acceptable when $\vect A$ is dense because the loop length (Line 4 of Alg.~\ref{algo:kernel}) is always $K$, and are no branches (Line 8 of Alg.~\ref{algo:kernel}). 
However, this approach has several disadvantages when $\vect A$ becomes sparse. 
Firstly, because each token corresponds to different rows of $\vect A$, the loop length of each thread are different (Line 4 of Alg.~\ref{algo:kernel}). In this case, all the threads need to wait until the longest loop is finished, introducing long waiting time. Secondly, there are branches in the sampling procedure, for example, the branch choosing which distribution to sample (Line 8 of Alg.~\ref{algo:kernel}). The branches cause\chris{alternatives for ``cause''?} the thread divergence problem, \ie, if some of the threads go to one branch and other threads go to another branch, the warp need to perform the instructions of both branches, which again increases the waiting time. Finally, the access to $\vect A$ is \emph{unconcealed} (Line 5 of Alg.~\ref{algo:kernel}) because the threads are accessing different and discontinuous addresses of the global memory.
\chris{Kaiwei: please make sure if what I said is correct}

\chris{TODO: double-check context of pre-processed sampling.}

To overcome the disadvantages of thread-based sampling, SaberLDA adopts \emph{warp-based sampling}, where all the threads in a warp collaboratively sample a single token. However, there are also challenges for warp-based sampling --- all the operations need to be vectorized to maximize the performance. 

We now present our vectorized implementation. As mentioned in Sec.~\ref{sec:pseudo-code}, the sampling involves computing the element-wise product (Line 4-7 of Alg.~\ref{algo:kernel}), randomly decide which branch to sample (Line 8 of Alg.~\ref{algo:kernel}), and depending on which branch is chosen, sample from $P$ (Line 9 of Alg.~\ref{algo:kernel}), or sample with the pre-processed tree (Line 11 of Alg.~\ref{algo:kernel}). Before starting the presentation, we also remind the readers that $\vect A$ is in global memory, and all the other data items such as $\vect B_v$ and $T_v$ are pre-loaded into shared memory taking advantage of PDOW as discussed in Sec.~\ref{sec:order-of-sampling}.

\subsubsection{Element-wise Product} 
The element-wise product step
is vectorizable by simply letting each thread process an index
and a warp compute the element-wise product for 32 indices
at a time. All the threads are efficiently utilized except for the
last 32 indices if the number of non-zero entries of Ad is not
a multiple of 32. The waste is very small since the number of
non-zero entries is typically much larger than 32, \eg, 100.

\subsubsection{Choosing the Branch} 
This step only consists a random number generation and a comparison, whose costs are negligible. Note that thread divergence will not occur as for the thread-based sampling since the whole warp goes to one branch or the other. 

\subsubsection{Sample from $P$} This step consists of the three steps mentioned in Sec.~\ref{sec:sampling-multinomial}, where the element-wise product is already computed. We need to generate a random number, and find its position in the prefix-sum array of $P$. 

Firstly, we need to vectorize the computation of the prefix sum, \ie, computing the prefix sum of 32 numbers using 32 threads. Because of the data dependency, the prefix sum cannot be computed with a single instruction. Instead, \textbf{\_\_shuf\_down} operations are required, and the prefix sum can be computed in $O(\log_2 32)$ instructions~\cite{prefixsum}. We refer this routine as \textbf{warp\_prefix\_sum}. \lkw{New paragraph here?}Given the prefix sum, we need to figure out the index of the first element which is greater than or equal to the value. This can be achieved in two steps:
\begin{enumerate}
    \item The warp-vote intrinsic \textbf{\_\_ballot} forms a 32-bit integer, where the $i$-th bit is one if the $i$-th prefix sum is greater than or equal to the given value~\cite{ballot},
    \item The \textbf{\_\_ffs} intrinsic returns the index of the first bit 1 of a 32-bit integer,
\end{enumerate}
where we refer these two steps as \textbf{warp\_vote}, which returns an index that is greater than or equal to the given value, or -1 if there is no such index. 

Deferring the discussion for the pre-processed sampling for a while, Fig.~\ref{fig:warp-sampling} is an example code of our vectorized warp-based sampling, where we omit some details such as whether the vector length is a multiple of the warp size, and the \textbf{warp\_copy(a, id)} function returns $a$ on the thread $id$. There is no waiting or thread divergence issue as discussed above. 

Besides the eliminated waiting time and thread divergence, the memory access behavior of our implementation is good as well. For the element-wise product, the access to $\vect A$ is continuous. More specifically, the warp accesses two 128-byte cache lines from the global memory, and each thread consumes two 32-bit numbers (an integer index and a float32 value). The accesses to $\vect B$ are random, but they are still efficient since the current row $\vect B_v$ is in shared memory. All the rest operations only access data items in the shared memory, \eg, $P$.

\subsubsection{$W$-ary Sampling Tree} 
\lstset{escapechar=@,style=customc}

\begin{figure}
\begin{lstlisting}[]
struct Wary_tree {
    float L1, L2;
    __shared__ Array L3, L4;
    
    Wary_tree(Array p)
    {
        L4.Alloc(p.size())
        L4 = array_prefix_sum(p);
        L3.Alloc(L4.size()/32)
        for (int i = thread_id; i<L3.size() ; i += 32)
            L3[i] = L4[i*32-1];
        L2 = L3[thread_id*32-1];
        L1 = warp_copy(L2, 31);
    }
    float Sum() { return L1; }
    int Sample(RandomSeed &seed)
    {
        float x = RandomFloat(seed) * L1;
        int off3 = warp_vote(L2 >= x) * 32;
        int off4 = (off3 + warp_vote(L3[off3+thread_id] >= x))*32;
        return off4 + warp_vote(L4[off4+thread_id]>=x);
    }
};
\end{lstlisting}
\caption{W-ary Sampling Tree.
\label{fig:w-ary}}
\end{figure}

We now present the deferred details of the pre-processed sampling (Line 11 of Alg.~\ref{algo:kernel}). The pre-processed sampling is new in sparsity-aware algorithms, since the vanilla algorithm does not break the original sampling problem into sub-problems. Therefore, previous GPU-based systems do not have this problem. 
As discussed in Sec.~\ref{sec:sampling-multinomial}, the pre-processed sampling problem is essentially the same problem as sampling from $P$, but there are only $V$ different sampling problems, so we can  pre-process for each problem to accelerate this process. 
In previous literature, there are two main data structures for that problem, and we briefly review them.
\begin{itemize}
    \item An alias table~\cite{walker1977efficient} can be built in $O(K)$ time, and each sample can be obtained in $O(1)$ time afterwards. However, building the alias table is sequential;
    \item A Fenwick tree~\cite{yu2015scalable} can be built in $O(K)$ time, and each sample can be obtained in $O(\log_2 K)$ time afterwards. However, the branching factor of the Fenwick tree is only two, so the 32-thread GPU warp cannot be fully utilized. 
\end{itemize}
Both approaches are designed for CPU, and are slow to construct on GPU because they cannot be vectorized. Vectorization is critical because using only one thread for pre-processing can be much slower than using a full warp for pre-processing. To allow vectorized pre-processing, we propose a \treename tree, which can be constructed in $O(K)$ time \emph{with full utilization of GPU warp}. Subsequent samples can be obtained in $O(\log_W K)$ time, where $W$ is the number of threads in a GPU warp, \ie, 32.

We emphasize that our main focus is on the efficient construction of the tree instead of efficient sampling using the tree, because the cost of  sampling using the tree is negligible comparing with sampling from $P$. Moreover, the sampling using our  \treename tree is efficient anyway because  $K$ is in the order of 10,000, and $\log_W K=4$, so $O(\log_W K)=O(4)$ is at the same level of the $O(1)$ alias table algorithm. 

\begin{figure}[t]
\centering
\includegraphics[width=\linewidth]{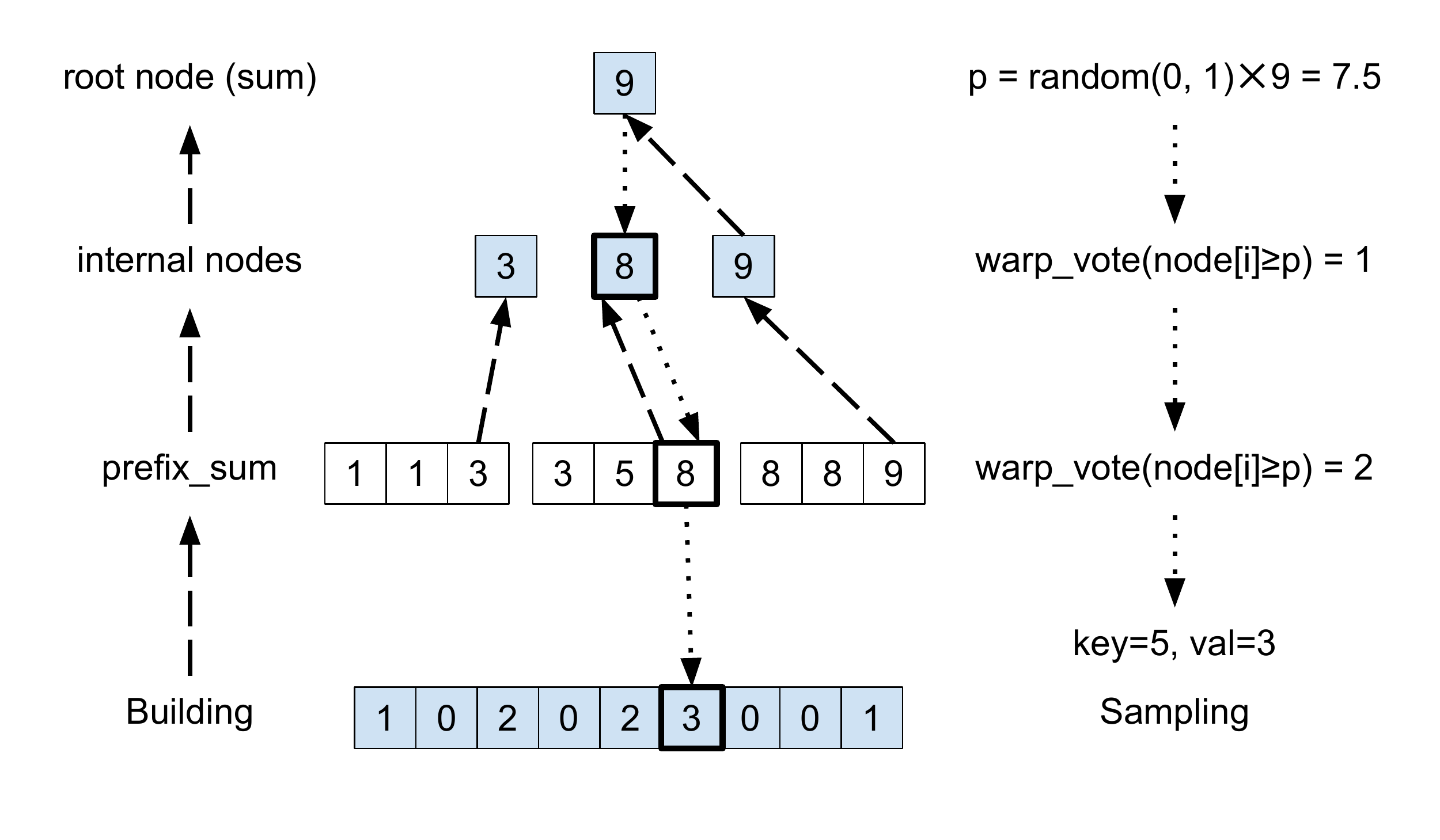}
\caption{Building a 3-ary tree
\label{fig:sampling1} and sampling from the tree.}
\end{figure}

Our \treename tree is designed for efficiently finding the location of a given number in the prefix-sum array. Each node of the tree stores a number, where the bottom-most level nodes store the prefix-sums of the given array. The length of an upper level is equal to the length of the lower level divided by $W$, and the $i$-th node in an upper level is equal to the $iW-1$ node in the lower level. 
Constructing the tree is illustrated in Fig.~\ref{fig:sampling1}. This procedure is efficient because all the nodes in one layer can be constructed in parallel. Therefore, the GPU warp can be efficiently utilized with the \textbf{warp\_prefix\_sum} function we mentioned before, which uses $W$ threads to compute the prefix-sum of $W$ numbers in parallel.

To find the position of a given value in the prefix-sum array, we recursively find the position at each level, from top to bottom (Fig.~\ref{fig:sampling1}). 
Based on the particular construction of our tree, if the position on the $l$-th level is $i$, then the position at the $l+1$-th level is between $iW$ and $iW+W-1$. Therefore, only $W$ nodes need to be checked on each level. This checking can be done efficiently using the \textbf{warp\_vote} function we mentioned before, and the memory access is efficient because the tree is stored in the shared memory (Sec.~\ref{sec:order-of-sampling}), and it only needs to read $W$ continuous floating point numbers, \ie, a 128-byte cache line for each level. 

The amount of memory accesses can be further reduced. We use a four-level tree for SaberLDA, which supports up to $W^3=32,768$ topics. The first and second layers contain only 1 and 32 nodes, respectively, and they can be stored in the thread registers. In this way, only two shared memory cache lines for level 3 and 4 are accessed per query. Fig.~\ref{fig:w-ary} is the code for the \treename tree.

\subsection{Efficient Count Matrix Update}\label{sec:ssc}

Finally, we discuss how to efficiently update the count matrices $\vect A$ and $\vect B$. When the sampling of all tokens with the same word is finished, the corresponding row $\vect B_v$ of the word-topic count matrix is ready to be updated. The \textbf{atomicAdd} function must be used because there may be multiple workers updating the same row, but the overhead is very low since the time complexity of updating is lower than the time complexity of sampling. 
The word-topic probability matrix $\vect{\hat{B}}$ can be easily generated according to Eq.~(\ref{eqn:preprocess}) from $\vect B$ after all the updates of the latter are finished. Maximal parallel performance can be achieved since both matrices are dense.

\begin{figure}
\centering
\includegraphics[width=\linewidth]{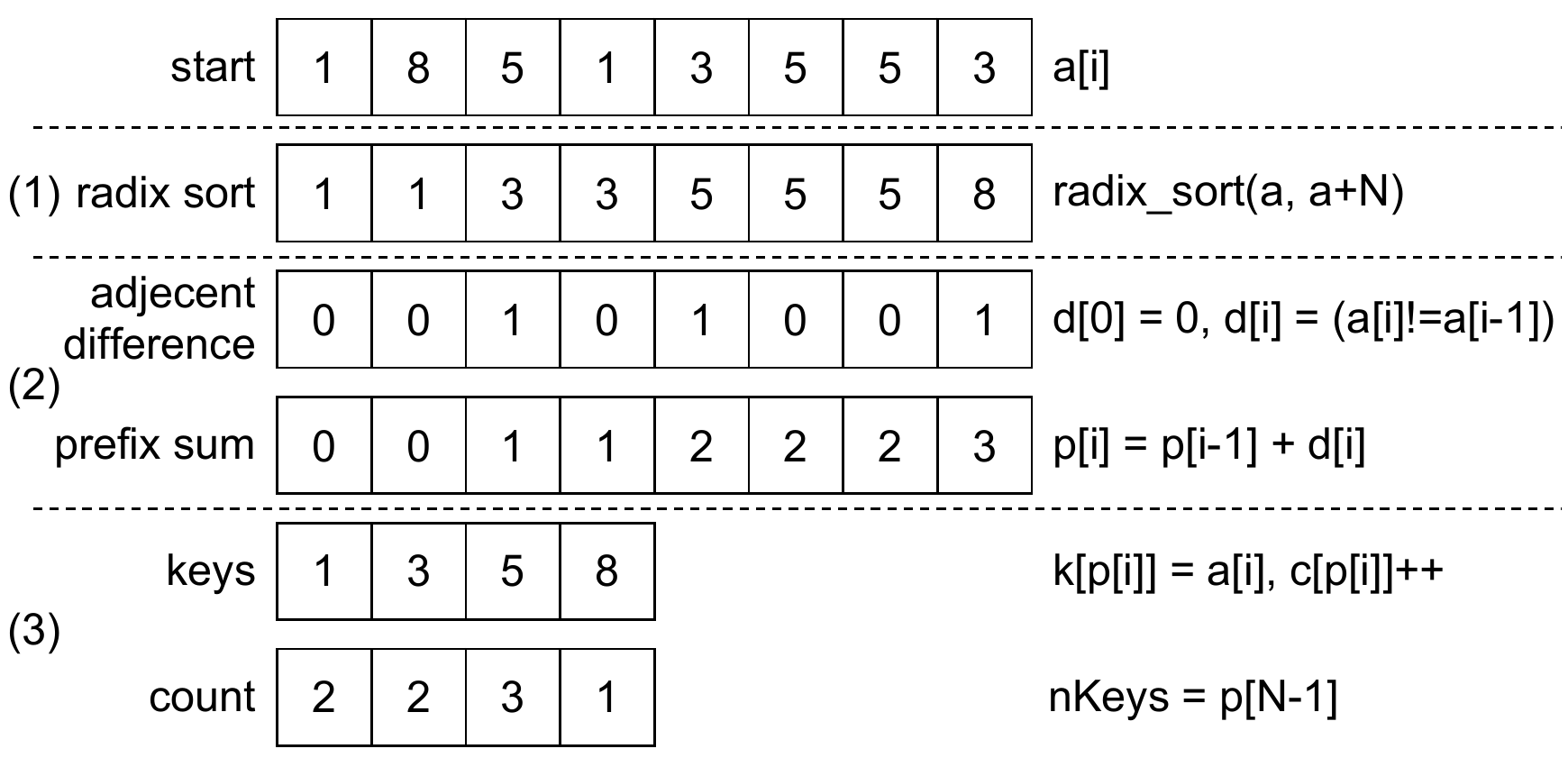}
\caption{An example for segmented count \chris{TODO indicate step1, 2 and 3 on the figure} \label{fig:counting}}
\end{figure}

However, the update of the document-topic matrix $\vect A$ is challenging since $\vect A$ is sparse. To update an entry of a sparse matrix, one must find the location of that entry, which is difficult to vectorize. Therefore, instead of updating, we \emph{rebuild} the count $\vect A$ after the sampling of each partition is finished. 

A na\"ive approach of rebuilding the count matrix is to sort all the tokens by first document-id $d$ then topic assignment $k$, and perform a linear scan afterwards. However, the sorting is expensive since it requires to access the global memory frequently. Moreover, the efficiency of sorting decreases as the chunk size increases. 


We propose a procedure called \emph{shuffle and segmented count}
(SSC) to address this problem. To rebuild the count matrix,
we first perform a \emph{shuffle} to organize the token list by the
document-id’s, \ie, segment the token list as smaller lists
where the tokens in each list share the same document-id.
Since the document-id’s of the tokens are fixed, the shuffle
can be accelerated with a pre-processed pointer array, which
points each token to its position in the shuffled list. Therefore,
to perform shuffling, we only need to place each token according
to the pointers. Furthermore, we can reduce the accesses
to global memory by creating the counts for each smaller token
list individually, where these token lists are small enough
to fit in the shared memory.

Creating the counts is a common problem as known as  \textit{segmented count}, which means for  \textit{segmenting} the tokens by $d$ and \textit{counting} $k$ in each segment. Unlike similar problems such as segmented sort~\cite{segsort} and segmented reduce~\cite{segreduce} which have fully  optimized algorithms for GPUs, efficient GPU solution of segmented count is not well studied yet. 

We propose a solution of segmented count which is sufficiently efficient for SaberLDA. Our procedure consists of three steps, as illustrated in Fig.~\ref{fig:counting}:
\begin{enumerate}
\item Perform a radix sort by $k$ in shared memory;
\item Calculate the prefix sum of the adjacent difference, in order to get the number of different topics and the order number of each topic;
\item Assign the topic number at corresponding order number, and increase the counter of the same topic.
\end{enumerate}


\subsection{Implementation Details and Load Balancing}
In SaberLDA, a word and a token are processed with a a block and a warp, respectively. Each block has its own shared memory to hold the rows $\vect{\hat{B}}_v$ and $\vect{B}_v$, where the former is fetched from the global memory before sampling any tokens in the current word $v$ and the latter is written back to global memory after the sampling. To minimize the number of memory transactions, we also align the memory address of each row of $\vect A$ by 128 bytes.

\chris{Please help revise the language in this part, I may write something inappropriate, for example, a block cannot process the data, the SM processes the data}
The workload for each block is determined by the size of its chunk, the workload for each block is determined by the number of tokens in the word, and the workload for each warp is determined by the number of non-zero entries in the document. To minimize the imbalance of the workload, we apply dynamic scheduling at all the three levels, \ie, a block fetches a new word to process when it is idle, and similarly, a warp fetches a new token to process when it is idle.


Because the term frequency of a natural corpus often follows
the power law~\cite{kingsley1932selective}, there are a few high-frequency
words which have much more tokens than other low-frequency
words. Therefore, the block level workload is more
imbalanced than the other two levels. To address this problem,
we further sort the words decreasingly by the number of
tokens in it, so that the words with most number of tokens
will be executed first, and then the words with small number
of tokens are executed to fill the imbalance gaps.

\section{Evaluation}\label{sec:eval}

\begin{table}[t]
\centering
\caption{Statistics of various datasets, where $T$ is the total number of words in the corpus.}\label{tbl:datasets}
\begin{tabular}{l | llllll}
\hline
Dataset & $D$ & $T$ & $V$  & $T/D$ & \\ \hline
NYTimes~ \cite{asuncion2007uci}          & 300K & 100M & 102k  & 332  & \\
PubMed~ \cite{asuncion2007uci}           & 8.2M & 738M & 141k  & 90     & \\
ClueWeb12 subset & 19.4M& 7.1B & 100k   & 365 & \\

\hline
\end{tabular}
\end{table}

In this section, we provide extensive experiments to analyze
the performance of SaberLDA, and compare SaberLDA with
other cutting-edge open source implementations on various
widely adopted datasets listed in Table~\ref{tbl:datasets}.

The code of SaberLDA is about 3,000 lines, written in CUDA and C++, and compiled with NVIDIA nvcc 8.0 and Intel ICC 2016. Our testing machine has two Intel E5-2670v3 CPUs, with 12 cores each CPU, 128 GB main memory and a NVIDIA GTX 1080 GPU. 

The hyper-parameters $\alpha = 50/K$ and $\beta = 0.01$ are set according to  previous works~\cite{yao2009efficient,li2014reducing,yu2015scalable,chen2016warplda}. 

The training time of LDA depends on both the number of iterations to converge and the time for each iteration. The former depends on the algorithm, \eg, variational Bayes algorithm~\cite{blei2003latent} typically requires fewer iterations than ESCA~\cite{zaheer2015exponential} to converge. The latter depends on the time complexity of sampling each token as well as the implementation, \eg, the sparsity-aware algorithm has $O(K_d)$ time complexity and performs faster than the $O(K)$ vanilla algorithms. Therefore, we use various metrics to compare LDA implementations:
\begin{itemize}
\item We use \emph{time per iteration} or \emph{throughput} to compare different
implementations of the same algorithm, \eg, compare
SaberLDA, which is a GPU implementation of the ESCA
algorithm, with a CPU implementation of the same algorithm,
because they require the same number of iterations
to converge. The throughput is defined as the number of
processed tokens divided by the running time, and the unit
is million tokens per second (Mtoken/s).

\item Since different algorithms require different numbers of iterations
to converge, the time per iteration metric is no
longer informative. Therefore, we compare different algorithms
by \emph{the required time to converge to a certain model quality}. The model quality is assessed by \emph{holdout log-likelihood} per token, using the partially-observed document
approach ~\cite{wallach2009evaluation}. Higher log-likelihood indicates better
model quality.

\end{itemize}




\subsection{Impact of Optimizations}\label{sec:decomposition}
\begin{figure}[]
\centering
\includegraphics[width=0.8\linewidth]{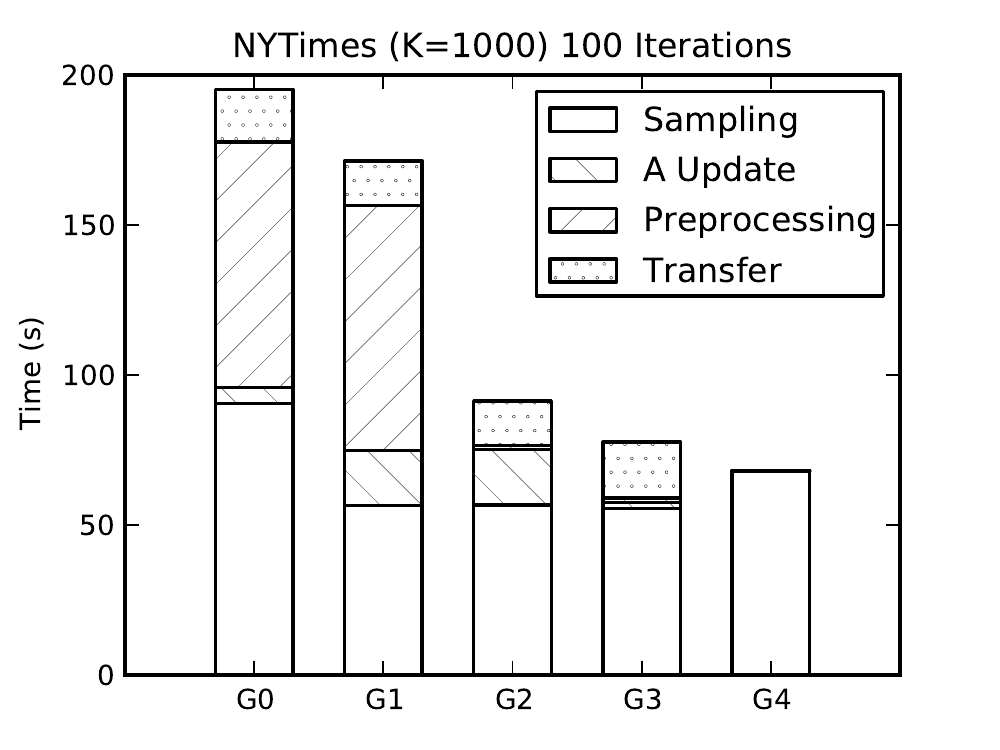}
\caption{Impact of optimizations.
\label{fig:optimize}
G0: Baseline; 
G1: PDOW;
G2: \treename tree;
G3: SSC;
G4: Asynchronous.}
\end{figure}

We first investigate the impact of each optimization technique we proposed in Sec.~\ref{sec:design-and-implementation}, by training LDA on the NYTimes dataset with 1,000 topics for 100 iterations. The results are shown in Fig.~\ref{fig:optimize}, where the total elapsed time is decomposed as the \textbf{Sample} function, rebuilding document-topic matrix $\vect A$, constructing pre-processed data structures for sampling (pre-processing), and data transferring between CPU and GPU.

G0 is the most straightforward sparsity-aware implementation
on GPU which sorts all tokens by documents, performs
the pre-processed sampling with the alias table, and builds
count matrices by na\"ive sorting of all tokens. G1 adopts the
PDOW strategy proposed in Sec.~\ref{sec:pdow}, and the time of sampling
is reduced by almost 40\% because of the improved locality
for sampling. Note rebuilding the doc-topic matrix A
takes more time in G1 than in G0, because the tokens are ordered
by word, and the sorting becomes slower.

\begin{figure*}[t]

\begin{minipage}[t]{0.31\linewidth}
\includegraphics[width=\linewidth]{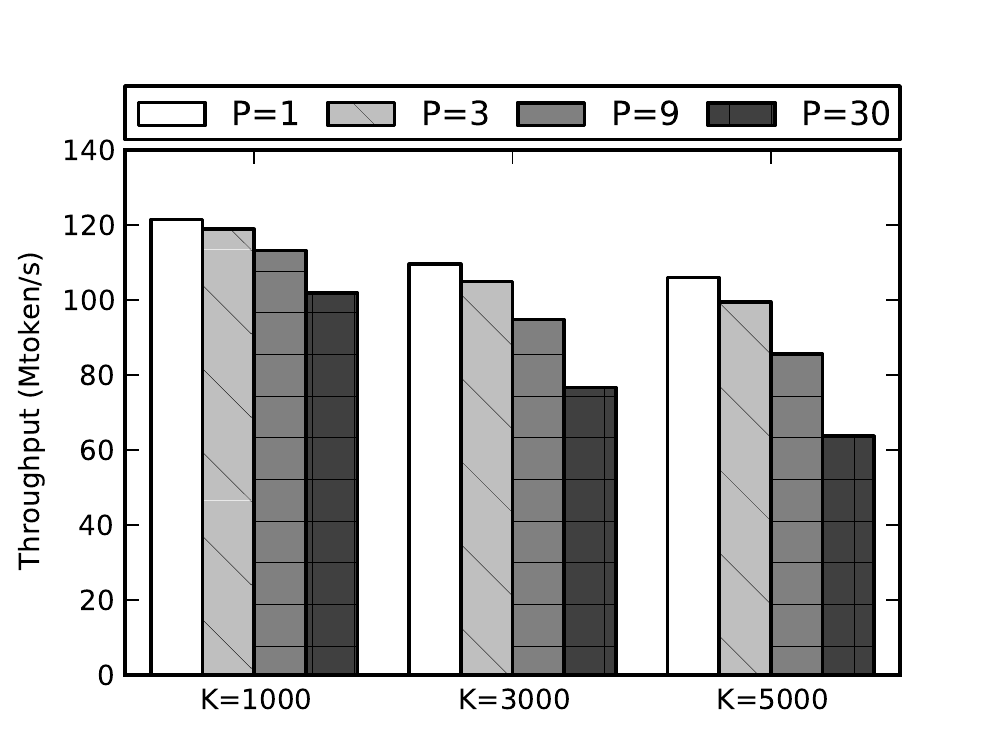}
\centering
(a)
\end{minipage}
\begin{minipage}[t]{0.31\linewidth}
\includegraphics[width=\linewidth]{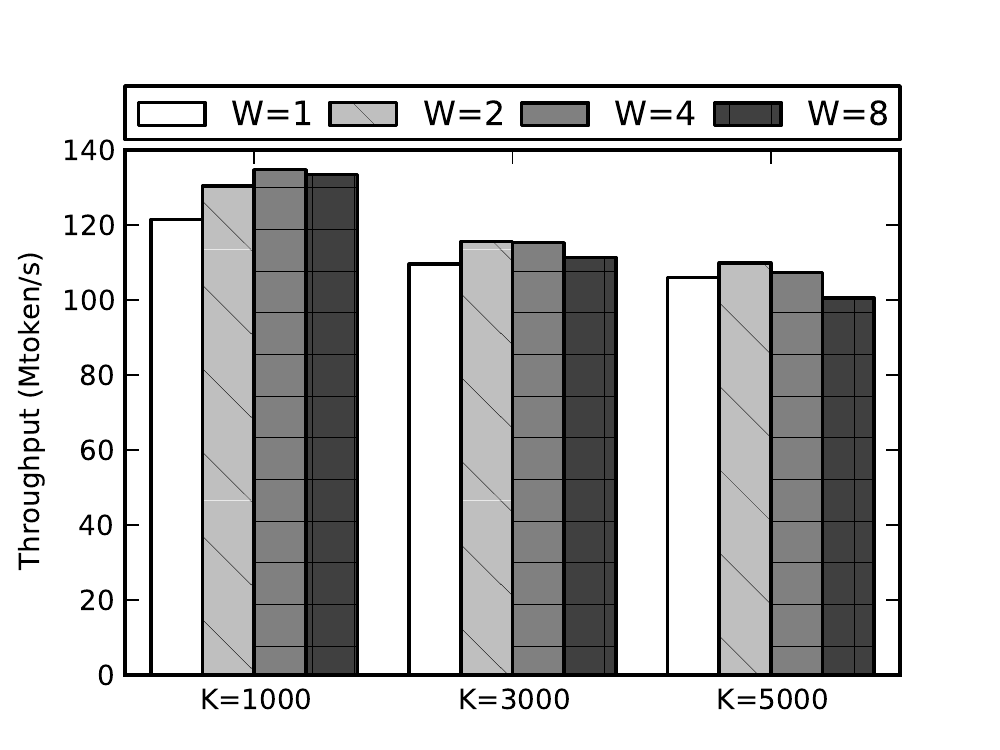}
\centering
(b)
\end{minipage}
\begin{minipage}[t]{0.31\linewidth}
\includegraphics[width=\linewidth]{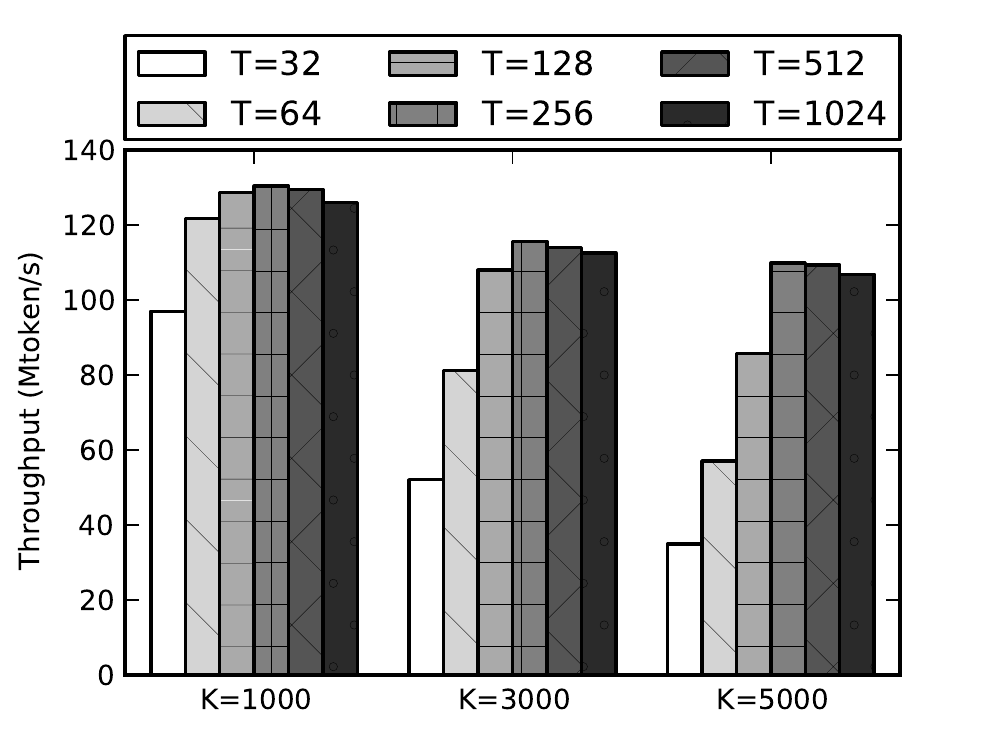}
\centering
(c)
\end{minipage}

\centering
\caption{\label{fig:tuning} Dateset: NYTimes. (a) Performance of different number of partitions. \chris{Reflect (a)}(b) Performance of different number of workers. (c) Performance of different number of threads. }
\end{figure*}

The bottleneck of G1 is the construction of the alias table, which is hard to vectorize. In G2, we replace the alias table with the \treename tree, which fully utilizes warps to greatly reduce the construction time by 98\%. 

Then, we optimize the rebuilding of the document-topic matrix $\vect A$ with SSC (Sec.~\ref{sec:ssc}), which reduces the rebuilding time by 89\%. Up to now, the time for updating $\vect A$ and pre-processing is neglectable.  

Finally, in G4, we enable multiple workers running asynchronously to hide the data transfer between CPU and GPU. It reduce 12.3\% of total running time in this case. Overall, all these optimizations achieve 2.9x speedup comparing with the baseline version G0. 
We emphasize that even G0 is already
highly optimized, and should handle more topics than previous
GPU implementations because it still adopts the sparsity-aware
algorithm whose time complexity is$ O(K_d)$.

\subsection{Performance Tuning}\label{sec:tuning}

\begin{table}[t]
\centering
\caption{Memory bandwidth utilization, NYTimes, $K=1000$.}
\label{tab:memory}
\begin{tabular}{l|ll}
\hline
                 & Throughput (GB/s) & Utilization \\ \hline
Global memory    & 144              & 50\%        \\
L2 cache         & 203              & 30\%        \\
L1 unified cache & 894              & 20\%        \\
Shared memory    & 458              & 20\%       \\ \hline
\end{tabular}
\end{table}



Tuning parameters, such as the number of workers, the number of chunks and the number of threads in a CUDA kernel can largely affect the performance. In order to fully understand the impact of tuning parameters, we analyze the performance of SaberLDA under different parameter settings.  We evaluate the total running time of the first 100 iterations on the NYTimes dataset with the number of topics varying from 1,000, to 3,000, and to 5,000.

\subsubsection{Number of Chunks}
Firstly, we analyze the single worker performance with various numbers of partitions, as shown in Figure~\ref{fig:tuning} (a). We can see that the performance degrades with more partitions because of the degraded locality, but partitioning is necessary when the dataset is too large to be held in GPU. We keep the number of partitions as small as possible to maximize the performance. 

\subsubsection{Number of Workers}
Figure~\ref{fig:tuning} (b) presents the performance with different numbers of workers. We fix the number of chunks to 10. Using multiple workers can hide the data transfer time between GPU and CPU, and reduce the overall time consumption. We observe a 10\% to 15\% speedup from single worker to 4 workers, where the speedup is quite close to the proportion of data transferring shown in Fig.~\ref{fig:optimize}. 

\subsubsection{Number of Threads}

Tuning the number of threads in each block maximizes the system performance. For the kernel function, more threads imply fewer active blocks, which reduces the total shared memory usage in a multiprocessor, but increases the in-block synchronization overhead of warps. Figure~\ref{fig:tuning} (c) shows that setting 256 threads in a block always achieves the best performance for various numbers of topics.

\subsection{Profiling Analysis}

Next, we analyze the utilization of hardware resources with NVIDIA visual profiler. We focus on memory bandwidth because LDA is a memory intensive task~\cite{chen2016warplda}.


Table~\ref{tab:memory} is the memory bandwidth utilization of the first 10
iterations on NYTimes with $K=1,000$. Statistics show that
the throughput accessing global memory reaches more than
140 GB/s, which is about 50\% of the bandwidth. This shows
clear advantage over CPUs, given that the bandwidth between
main memory and CPU is only 40 to 80 GB/s. The throughput
of the L2 cache, unified L1 cache and shared memory is
203GB/s, 894GB/s and 458GB/s respectively, while the utilization
is lower than 25\%. Therefore, these are not the bottleneck
of the overall system performance.

We further use performance counters to examine of kernel
function. The memory dependency is the main reason of instruction
stall (47\%), and the second reason is execution dependency
(27\%). The hotspot is computing the element-wise
product between sparse vector $\vect A_d$ and dense vector $\vect{\hat{B}}_w$, which
is expected because it accesses global memory.


\subsection{Comparing with Other Implementations}

\begin{figure*}[t]
\centering
\begin{minipage}[t]{0.665\linewidth}
\includegraphics[width=\linewidth]{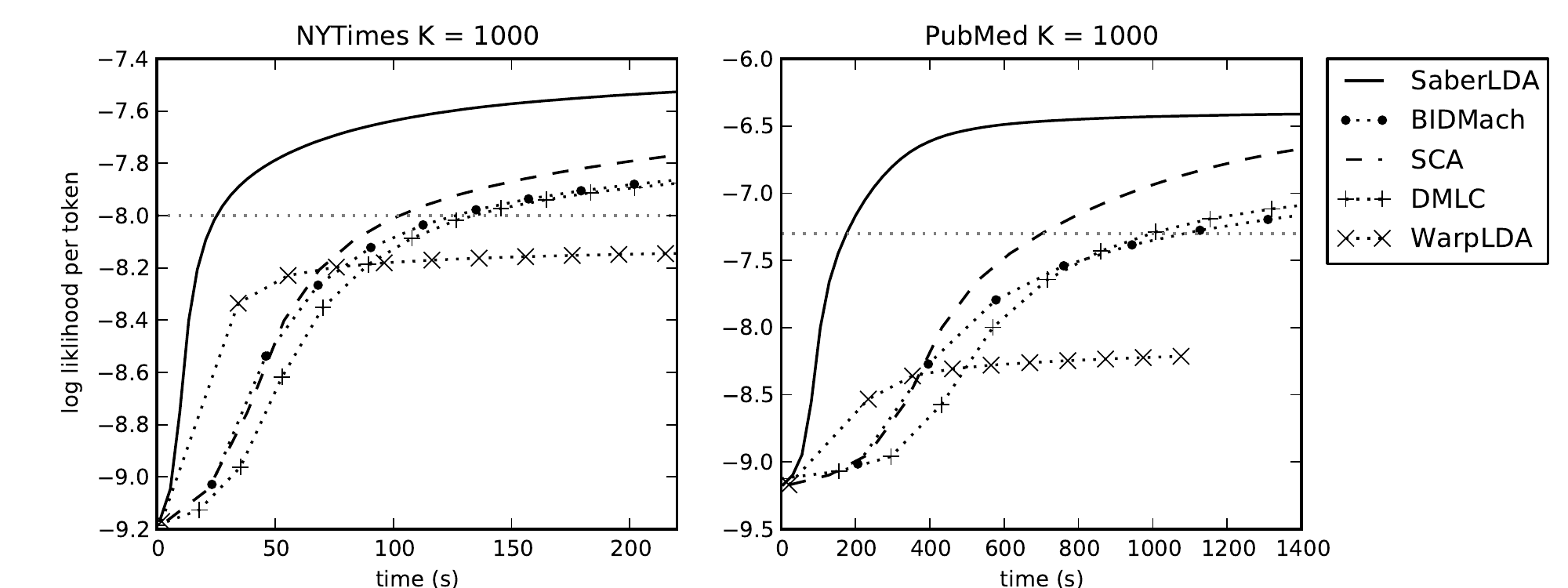}
\caption{Convergence over time with 1000 topics. The left is evaluated
on the NYTimes dataset, while the right is on the PubMed dataset.\label{fig:pvt}}
\end{minipage}
\hfill
\begin{minipage}[t]{0.33\linewidth}
\includegraphics[width=\linewidth]{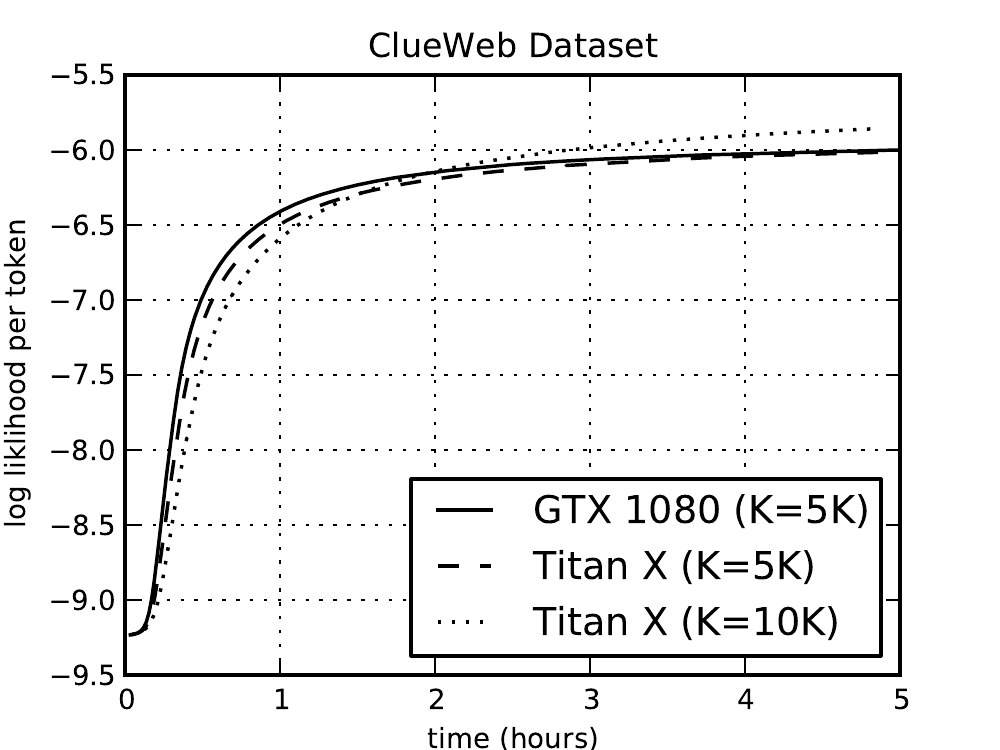}
\caption{The convergence of SaberLDA on ClueWeb Subset. \label{fig:huge}}
\end{minipage}
\end{figure*}

We compare SaberLDA with a state-of-the-art GPU-based implementation
BIDMach~\cite{zhao2015same} as well as three state-of-the-art
CPU-based methods, including ESCA (CPU), DMLC~\cite{dmlclda}
and WarpLDA~\cite{chen2016warplda}. BIDMach~\cite{zhao2015same} is the only open-sourced
GPU-based method. BIDMach reports better performance
than Yan et. al’s mehtod~\cite{zhao2015same,yan2009parallel}, and Steele and Tristan’s
method only reports tens of topics in their paper. Therefore,
BIDMach is a strong competitor. ESCA (CPU) is a carefully
optimized CPU version of the ESCA algorithm which
SaberLDA also adopts. DMLC has various multi-thread
LDA algorithms on CPU, and we choose its best performing
FTreeLDA. WarpLDA is a state-of-art distributed LDA
implementation on CPU based on the Metropolis-Hastings algorithm,
and uses a cache-efficient $O(1)$ sampling algorithm
to obtain high per-iteration throughput.

We compare the time to converge of these implementations
on NYTimes and PubMed datasets, with the number of topics
$K=1,000$. Figure~\ref{fig:pvt} shows the convergence over time. We
compare the time to converge to the per-token log-likelihood
of $-8.0$ and $-7.3$, for NYTimes and PubMed, respectively.
SaberLDA is about 5.6 times faster than BIDMach. We also
attempt to perform the comparison with 3,000 and 5,000 topics,
and find that BIDMach is more than 10 times slower than
SaberLDA with 3,000 topics, and reports an out-of-memory
error with 5,000 topics. This is as expected because the time
consumption of BIDMach grows linearly with respect to the
number of topics, and its dense matrix format is much more
memory consuming than SaberLDA.

Finally, SaberLDA is about 4 times faster than ESCA
(CPU) and 5.4 times faster than DMLC on the two datasets
with $K=1,000$, where WarpLDA converges to a worse local
optimum possibly because of its inexact Metropolis-Hastings
step and the different metric with its paperr~\cite{chen2016warplda} which we use
to assess model quality. This shows that SaberLDA is more
efficient than CPU-based implementations.

\subsection{A Large Dataset}
Finally, to demonstrate the ability of SaberLDA to process large datasets,  we test the performance of SaberLDA on a large subset of the ClueWeb dataset, which is a crawl of webpages.~\footnote{\url{http://www.lemurproject.org/clueweb12.php/}} We filter out the stop words and keep the remaining  100,000 most frequent words as our vocabulary, which is comparable to the vocabulary size of NYTimes. We use the entire CPU memory to hold as many documents as possible, which is  19.4 million, and the total number of tokens is 7.1 billion, which is about 10 times larger than the PubMed dataset. 

In this experiment, we also use GTX Titan X (Maxwell), which has 12GB global memory besides GTX 1080, which has 8GB global memory. We compare the performance of GTX 1080 and Titan X with 5,000 topics.


The algorithm convergences in 5 hours on both cards, where the throughput of GTX 1080 is 135 Mtoken/s and the throughput of Titan X is 116 Mtoken/s. With 10,000 tokens, it also convergences in 5
hours with a throughput of 92 Mtoken/s.

\section{Conclusions}\label{sec:conclusions}
We present SaberLDA, a high performance sparsity-aware
LDA system on GPU. Adopting sparsity-aware algorithms,
SaberLDA overcomes the problem of previous GPU-based
systems, which support only a small number of topics. We
propose novel data layout, warp-based sampling kernel, and
efficient sparse count matrix updating algorithm to address
the challenges induced by sparsity, and demonstrate the
power of SaberLDA with extensive experiments. It can efficiently
handle large-scale datasets with up to 7 billion tokens
and learn large LDA models with up to 10,000 topics, which
are out of reach for the existing GPU-based LDA systems.

In the future, we plan to extend SaberLDA to multiple
GPUs and machines. Developing algorithms that converge
faster and enjoy better locality is also our future work.

\bibliographystyle{plain}
\bibliography{main}




\end{document}